# Spatiotemporal THz emission from radial and longitudinal wakefields by co-propagating chirped lasers in magnetized rippled plasma

A. A. Molavi Choobini and F. M. Aghamir*
Dept. of Physics, University of Tehran, Tehran 14399-55961, Iran.

**Abstract:**

The excitation of radial and longitudinal wake-fields by two co-propagating chirped laser pulses in a rippled, magnetized plasma has been examined. This study aimed to clarify the spatiotemporal evolution of wake structures and assess their role in the generation of THz radiation. A Fourier-Bessel Particle-In-Cell (FBPIC) simulation framework, optimized for cylindrical geometries, has been employed to model the relativistic dynamics of plasma electrons under the combined influence of laser-induced ponderomotive forces and an external magnetic field. It has been shown that the beat frequency between the pulses modulates the ponderomotive force, driving nonlinear wake-field structures sustained by electron oscillations. Simulations performed with high spatial resolution have revealed that wake-field amplitude and coherence are strongly influenced by laser chirp, pulse duration, and plasma density. Distinct THz peaks have been identified in the Fourier-transformed spectra, with their amplitudes enhanced by resonant coupling between wake-field harmonics and the laser frequency modulation. Moreover, electron motion has been confined by the magnetic field, leading to improved energy gain and shaping of angular radiation patterns. These findings suggest that tailored laser and plasma configurations can be used to optimize energy transfer mechanisms, paving the way for more efficient wake-field usage and THz generation.

*Corresponding author: E-mail address: aghamir@ut.ac.ir

## I. Introduction

The interaction of intense laser pulses with plasma can excite plasma wake-fields—collective charge-density perturbations that follow the laser front. These wake-fields are capable of accelerating charged particles to relativistic energies over short distances, marking a significant advancement in compact accelerator technologies [1–5]. Beyond their role in particle acceleration, plasma wakefields have also been explored as a promising mechanism for generation of Terahertz (THz) radiation [6, 7]. THz emission arises from the nonlinear plasma response to ultrashort, intense laser pulses. As these pulses propagate through the plasma, they induce charge separation and transient currents that emit broadband electromagnetic radiation in the THz range. This process offers a compact, tunable, and high-power alternative to conventional THz sources, with potential applications in imaging, spectroscopy, and ultrafast diagnostics [8, 9]. To enhance THz yield, schemes employing multiple laser pulses co-propagating or counter-propagating, chirped or modulated--have been employed, while external magnetic fields provide additional control over wave polarization, coherence, and directionality [10–16].

Extensive theoretical and experimental efforts have examined strategies for optimizing THz generation in laser–plasma interaction systems. Initial studies by Wang et al. [17] highlighted the synergy between laser pulses and electron bunches in plasma-based particle acceleration. Subsequent studies explored density gradients [18], plasma gratings [19] and rippled density profiles [20, 21] as mechanisms for phase matching and wake-field enhancement. Chirped laser pulses have been shown to significantly affect wake-field dynamics, where positive chirps can amplify wake amplitudes and improve electron trapping, while negative chirps suppress them [22-24]. These studies establish chirp control as a vital tuning mechanism in laser wake-field acceleration (LWFA) and THz generation. Other approaches exploited nonlinear plasma effects, ionization fronts, and high-current electron bunches to generate narrowband or single-cycle THz pulses [25-27]. Magnetized plasma environments provide additional control over emission characteristics. Tailliez et al. [28] used PIC simulations to show that a strong axial magnetic field introduces circular polarization in the emitted THz radiation. External fields can enhance efficiency, introduce polarization control, and improve coherence [29-31]. Collectively, these studies underscore the sensitivity of THz generation to laser and plasma parameters and highlight the need for tunable schemes that combine chirp control, plasma structuring, and magnetic confinement [32, 33]. These studies provide insights into the parameter sensitivity and coherence requirements for efficient THz generation.

A novel mechanism has been introduced for the generation of both radial and axial wake-fields in the THz frequency range by employing two co-propagating chirped laser pulses in a rippled, magnetized plasma channel. This approach exploits the temporal frequency modulation of chirped pulses to dynamically control the ponderomotive force and the evolution of wakefields. The presence of a magnetic field enhances the process by stabilizing and amplifying the wake structures through resonant interactions among the plasma frequency, cyclotron motion, and the modulated laser field. The dynamics are modeled using the Fourier-Bessel Particle-In-Cell (FBPIC) framework in cylindrical geometry, providing efficient and accurate simulation of laser–plasma interactions. The wake-field formation, nonlinear wave–particle coupling, and beam–plasma instabilities have successfully been resolved. The influence of key parameters—laser chirp, pulse length, plasma density, and magnetic field strength—has been systematically investigated. An increase in the chirp parameter intensifies the initial field amplitudes and sharpens oscillatory features, indicating stronger laser–plasma coupling. Meanwhile, stronger magnetic fields confine electron motion and enhance longitudinal coherence, thereby improving energy transfer efficiency. This tunability has been thoroughly explored through theoretical modeling and PIC simulations, leading to the identification of optimal conditions for maximizing wakefield generation, as evidenced by the spectral features observed in the Fourier-transformed field distributions. In addition, It has been demonstrated that Gaussian laser profiles enhance both the spatial coherence and the strength of the wakefields. The dynamic synchronization via cross-phase modulation and plasma-induced group velocity matching improves energy transfer and THz emission. The interplay between coherence degradation—caused by collisional and nonlinear effects—and the temporal evolution of the wakefield has revealed novel pathways for maintaining coherence over longer propagation distances. The article is organized as follows: Section II presents the theoretical model for generating radial and axial wakefields in the THz frequency domain. Section III discusses the results, focusing on the characteristics of both longitudinal and transverse wakefields. Finally, Section IV provides the concluding remarks.

## II. Theoretical Model

Consider two chirped laser pulses propagating simultaneously through a rippled plasma in the positive $z$-direction. The pulses have angular frequencies $\omega_1$ and $\omega_2$, and wave numbers $k_1$ and $k_2$, respectively. Both pulses are linearly polarized along the $x$-direction. The electric field of combined laser pulses can be written as:

$$\vec{E}(z,t) = [E_{01}Cos(k_1 z - \omega_1 t) + E_{02}Cos(k_2 z - \omega_2 t)]\hat{e}_x \tag{1}$$

where the field amplitudes are defined as:

$$E_{0s} = E_{rs}\, e^{-\frac{r^2}{r_{0s}^2}}\, Sin\frac{\pi\xi}{L_s} e^{-\frac{t^2}{\tau_L^2}} \tag{2}$$

here $E_{rs}$ denotes the peak electric field, $r_{0s}$ the spot size of pulse, $\tau_L$ pulse duration, and $L_s$ the pulse length. The independent variable $\xi$ is given by $\xi = z - ct$, and the index $s = 1,2$ corresponds to each pulse. For axial propagation, the radial distance $r$ from the beam center is assumed to be zero, implying that the analysis is restricted to the laser pulse behavior along its central axis, ignoring radial variations. The laser frequencies undergo a linear chirp over time, described by:

$$\omega_s = \omega_{0s}(1 + \alpha c \omega_{0s}\xi) \tag{3}$$

Here $\xi = z - ct$ denotes the co-moving coordinate in the speed-of-light frame (where the laser envelopes appear stationary for slow evolution), while t (or $\tau$ in some notations) is the slow laboratory time characterizing driver evolution. The quasi-static approximation assumes $\partial/\partial t \ll \omega_p$ (or even beat frequency) for envelope variations, allowing neglect of explicit time derivatives in the plasma fluid equations except for fast oscillatory terms retained in the velocity response. Furthermore, $\omega_{0s}$ is the central frequency of the lasers, $\alpha$ is the frequency chirp parameter that determines the magnitude of frequency variation, and $c$ is the speed of light. In the absence of chirp ($\alpha = 0$), the laser frequencies remain constant. The central frequencies $\omega_{0s}$ are chosen such that the beat frequency, defined as the frequency difference between the two laser pulses $\Delta\omega$ is close to the plasma frequency $\omega_p$, enabling resonant excitation of plasma waves. In this beat-wave mechanism, two laser pulses modulate the ponderomotive force at the plasma frequency, unlike single-pulse schemes, allowing tunable wake-field generation through chirp control. Because the ponderomotive force is proportional to the gradient of the combined field intensity $|E_1 + E_2|^2$, this modulation resonantly drives plasma oscillations. The resulting nonlinear interaction amplifies electron density fluctuations and produces spatial gradients in both the axial ($z$) and radial ($r$) directions, which generate the corresponding ponderomotive forces. These forces, in turn, induce electron density perturbations that give rise to the longitudinal wake-field and transverse (radial) field components of the plasma wave. The laser field envelopes in Eqs. (1)–(3) are assumed to propagate in the underdense plasma with negligible dispersion over the interaction length, consistent with the quasi-static approximation where the laser evolution is slow compared to the plasma response time. This neglects plasma-induced group velocity dispersion, frequency up/down-shifting, and relativistic self-focusing effects on the envelopes themselves, which are instead captured self-consistently in the FBPIC simulations. Spatial effects such as relativistic self-focusing are not included in the analytical fluid model but influence pulse propagation and wake excitation in the numerical results. To suppress phenomena such as slipping instability or electromagnetic filamentation, an external static magnetic field $\vec{B}_{ext} = B_0\hat{e}_z$ is applied along the propagation direction. The magnetic modulation wavelength is assumed to be much larger than the wake-field wavelength, ensuring the validity of the slowly varying field approximation. The temporal evolution of the electron velocity and density can be evaluated under these conditions:

$$\frac{\partial \vec{v}}{\partial t} + \nu_c \vec{v} = \frac{-e}{m_e \gamma}\left[\vec{E} + \frac{1}{c}\left(\vec{v} \times \vec{B}\right)\right] - \frac{1}{2}\vec{\nabla}(\vec{v}.\vec{v}) + \vec{v} \times \left(\vec{\nabla} \times \vec{v}\right) \tag{4}$$

and

$$\frac{\partial n}{\partial t} + \vec{\nabla}.(n\vec{v}) = 0 \tag{5}$$

here $e$ and $m_e$ denote the electron charge and mass, respectively. The collision frequency $\nu c$ accounts for energy dissipation and damping effects in plasma. In the parameter regime considered here -- where laser intensities are below the strongly relativistic threshold -- the relativistic factor $\gamma$ is set to unity, consistent with the quasi-static approximation. The electron density is expressed as $n = n_u + n_{Pe}$, where $n_u$ and $n_{Pe}$ represent the un-perturbed and perturbed components, respectively. Presuming the perturbed density changes as $n_{Pe} = n_0 e^{i(k_2-k_1)\xi} = n_0 e^{ik_T\xi}$, and using Poisson equation, the differential equation for time-varying first-order density perturbation can be written as:

$$\frac{\partial^2 n_{Pe}}{\partial t^2} + \nu_c \frac{\partial n_{Pe}}{\partial t} + n_{Pe}\left[\omega_p^2 + \omega_c^2 k_T^2 c^2 \left(\frac{1}{\omega_1^2 - \omega_c^2} + \frac{1}{\omega_2^2 - \omega_c^2}\right)\right]$$

$$= \frac{eE_{01}k_1 n_u}{m_e}\left[\frac{\omega_1^2}{\omega_1^2 - \omega_c^2} Sin(k_1 z - \omega_1 t) - \frac{\omega_1 \omega_c}{\omega_1^2 - \omega_c^2} Cos(k_1 z - \omega_1 t)\right]$$

$$+ \frac{eE_{02}k_2 n_u}{m_e}\left[\frac{\omega_2^2}{\omega_2^2 - \omega_c^2} Sin(k_2 z - \omega_2 t) - \frac{\omega_2 \omega_c}{\omega_2^2 - \omega_c^2} Cos(k_2 z - \omega_2 t)\right] \tag{6}$$

here $\omega_p$ is the plasma frequency, $K_T = k_2 - k_1$ the beat wave number, and $\omega_c = eB_0/m_e c$ the cyclotron frequency of plasma electrons in a collisionless-plasma. Simultaneous solution of Equations. 5 and 6 yields the first-order perturbed particle density:

$$n_{Pe} = \frac{eE_{01}k_1 n_u}{m_e \Pi_1}\left[\frac{\omega_1 \omega_c}{\omega_1^2 - \omega_c^2 + i\nu_c\omega_1} Cos(k_1 z - \omega_1 t) - \frac{\omega_1^2}{\omega_1^2 - \omega_c^2 + i\nu_c\omega_1} Sin(k_1 z - \omega_1 t)\right]$$

$$+\frac{eE_{02}k_2n_u}{m_e\Pi_2}\left[\frac{\omega_2\omega_c}{\omega_2^2-\omega_c^2+i\nu_c\omega_2}Cos(k_2z-\omega_2t)-\frac{\omega_2^2}{\omega_2^2-\omega_c^2+i\nu_c\omega_2}Sin(k_2z-\omega_2t)\right] \quad (7)$$

where $\Pi_1=\omega_p^2-\omega_1^2+i\nu_c\omega_1+\frac{2\omega_c^2k_1^2}{\omega_1^2-\omega_c^2+i\nu_c\omega_1}$ and $\Pi_2=\omega_p^2-\omega_2^2+i\nu_c\omega_2+\frac{2\omega_c^2k_2^2}{\omega_2^2-\omega_c^2+i\nu_c\omega_2}$. It is assumed that the electrons initially have zero velocity and are influenced solely by the combined electric field of the laser pulses. As two chirped laser pulses propagate through the magnetized plasma, the ponderomotive force of the wave field drives plasma electrons to oscillate at the laser frequencies. These rapid oscillations in charge density generate a longitudinal electric field and an associated electron velocity, expressed as:

$$E_{z,Long}=-E_{01}\Pi_1\omega_p^2\left[\frac{\omega_1^2}{\omega_1^2-\omega_c^2+i\nu_c\omega_1}Cos(k_1z-\omega_1t)+\frac{\omega_1\omega_c}{\omega_1^2-\omega_c^2+i\nu_c\omega_1}Sin(k_1z-\omega_1t)\right]$$

$$-E_{02}\Pi_2\omega_p^2\left[\frac{\omega_2^2}{\omega_2^2-\omega_c^2+i\nu_c\omega_2}Cos(k_2z-\omega_2t)+\frac{\omega_2\omega_c}{\omega_2^2-\omega_c^2+i\nu_c\omega_2}Sin(k_2z-\omega_2t)\right] \quad (8)$$

and

$$v_{z,Long}=\frac{eE_{01}\omega_p^2}{m_e\omega_1}\Pi_1\left[\frac{\omega_1\omega_c}{\omega_1^2-\omega_c^2+i\nu_c\omega_1}Cos(k_1z-\omega_1t)-\frac{\omega_1^2}{\omega_1^2-\omega_c^2+i\nu_c\omega_1}Sin(k_1z-\omega_1t)\right]$$

$$+\frac{eE_{02}\omega_p^2}{m_e\omega_2}\Pi_2\left[\frac{\omega_2\omega_c}{\omega_2^2-\omega_c^2+i\nu_c\omega_2}Cos(k_2z-\omega_2t)-\frac{\omega_2^2}{\omega_2^2-\omega_c^2+i\nu_c\omega_2}Sin(k_2z-\omega_2t)\right] \quad (9)$$

Since the laser frequency is much higher than the plasma frequency, the plasma electrons cannot collectively respond to the rapid oscillations of the wave. Since the laser frequencies $\omega_1$ and $\omega_2$ are much greater than the plasma frequency, the plasma electrons respond rapidly to the individual laser oscillations, leading to quiver velocities oscillating at $\omega_1$ and $\omega_2$ (as retained in Eq. (4) for electron velocity). However, the slow beat frequency $\Delta\omega\approx\omega_p$ drives the ponderomotive force modulation responsible for wake excitation. In the quasi-static approximation, the laser envelopes and wakefields evolve slowly on the time scale $\tau_e nv \gg 2\pi/\omega_p$ (or driver evolution time), while the fast optical cycles ($\ll 2\pi/\omega_p$) are averaged or retained only for quiver motion. Thus, explicit time derivatives $\partial/\partial t$ associated with envelope variations are neglected in the plasma fluid equations (consistent with the slowly varying envelope approximation), but fast oscillatory terms in velocity and current are kept computing the nonlinear ponderomotive drive. The plasma fluid equations are therefore solved on timescales larger than the plasma wave period (for wake evolution), which is much longer than the optical cycle but comparable to or longer than the pulse duration in resonant beat-wave regimes. Under these assumptions, the time-dependent Maxwell's equations are employed to describe wake-field generation. Considering a radial Gaussian profile for the laser pulses (as given in Eq. 2) and assuming axisymmetric field generation, Maxwell's equations are written in cylindrical-polar coordinates as:

$$\frac{\partial E_r}{\partial\xi}-\frac{\partial B_\theta}{\partial\xi}=\frac{4\pi}{c}J_r \quad (10a)$$

$$\frac{\partial E_r}{\partial\xi}-\frac{\partial E_z}{\partial r}=\frac{\partial B_\theta}{\partial\xi} \quad (10b)$$

$$\frac{\partial E_r}{\partial \xi} + \frac{1}{r}\frac{\partial (rB_\theta)}{\partial r} = \frac{4\pi}{c} J_z \tag{10c}$$

where $J_z$ and $J_r$ are the axial and radial (transverse) current densities, respectively. By substituting the quantities from Eqs. 8 and 9 into the Lorentz force equation, the differential equations for radial and axial velocity components can be derived:

$$\frac{\partial v_z}{\partial \xi} = \frac{e}{m_e c} E_z + \frac{\partial}{\partial \xi}\left(\frac{c}{4}\sum_{s=1}^{2} \Pi_s^2 \omega_p^2 a_{0s}^2 \frac{\omega_s^4 + \omega_s^2 \omega_c^2}{\left(\omega_s^2 - \omega_c^2 + i v_c \omega_s\right)^2}\right) \tag{11}$$

and

$$\frac{\partial v_r}{\partial \xi} = \frac{e}{m_e c} E_r - \frac{c}{r_0^2}\sum_{s=1}^{2} \Pi_s^2 \omega_p^2 r a_{0s}^2 \frac{\omega_s^4 + \omega_s^2 \omega_c^2}{\left(\omega_s^2 - \omega_c^2 + i v_c \omega_s\right)^2} \tag{12}$$

here $E_r$ and $E_z$ denote the radial and axial components of electric wake-fields. To derive equations 9 and 10, the longitudinal component of the gradient of the ponderomotive force ($\vec{\nabla}_z$) must be considered. This component acts on the spatial variations of the plasma particle velocity along the propagation direction, displacing electrons longitudinally and producing a density perturbation behind the laser pulse—referred to as the longitudinal wake-field. In contrast, the transverse component of the gradient ($\vec{\nabla}_\perp$) acts on the spatial variations of the particle velocity in the transverse direction. These variations arise from the non-uniform transverse intensity profile of the laser beam, which generates a transverse ponderomotive force. This force pushes electrons sideways, leading to transverse charge redistribution and the formation of transverse wake-fields. By combining Eqs. (10) – (12), two differential equations governing the axial and radial wake-field variations are obtained:

$$\left[\frac{\partial^2}{\partial \xi^2} + k_p^2\left(1 + \frac{n_{pe}}{n_u}\right)\right] E_z = -\frac{\partial}{\partial \xi}\left(\frac{1}{r}\frac{\partial (rB_\theta)}{\partial r}\right) - \frac{m_e c^2}{4e}\sum_{s=1}^{2} \Pi_s^2 \omega_p^2 \frac{\omega_s^4 + \omega_s^2 \omega_c^2}{\left(\omega_s^2 - \omega_c^2 + i v_c \omega_s\right)^2}\frac{\partial a_{0s}^2}{\partial \xi} \tag{13}$$

and

$$E_r = -\frac{1}{k_p^2}\frac{\partial^2 E_z}{\partial \xi \partial r} + \frac{m_e c^2}{e r_0^2}\sum_{s=1}^{2} \Pi_s^2 \omega_p^2 \frac{\omega_s^4 + \omega_s^2 \omega_c^2}{\left(\omega_s^2 - \omega_c^2 + i v_c \omega_s\right)^2} r a_{0s}^2 \tag{14}$$

where $k_p = \omega_p / c$ is the plasma wave number and $a_{0s} = eE_{0s}/m_e \omega_s$. By perturbatively expanding Eqs. (13) and (14), matching terms of equal order, and assuming a Gaussian pulse profile, the lowest-order solutions for the radial and axial wake-fields are obtained as:

$$E_z^{(1)} = \frac{m_e k_p c^2}{8e}\left(\sum_{s=1}^{2} \frac{\mathrm{a}_{rs}^2 \Pi_s^2 \omega_p^2 r}{\left(1 - \frac{k_p^2 L_s^2}{4\pi^2}\right)} \frac{\omega_s^4 + \omega_s^2 \omega_c^2}{\left(\omega_s^2 - \omega_c^2 + i v_c \omega_s\right)^2}\right)\left[Sin k_p (L - \xi) + Sin k_p \xi\right] \tag{15}$$

$$E_r^{(1)} = \frac{m_e c^2}{2 e r_0^2}\left(\sum_{s=1}^{2} \frac{\mathrm{a}_{rs}^2 \Pi_s^2 \omega_p^2 r}{\left(1 - \frac{k_p^2 L_s^2}{4\pi^2}\right)} \frac{\omega_s^4 + \omega_s^2 \omega_c^2}{\left(\omega_s^2 - \omega_c^2 + i v_c \omega_s\right)^2}\right)\left[Cos k_p \xi - Cos k_p (L - \xi) - Cos \frac{2\pi}{L} + 1\right] \tag{16}$$

In addition, The lowest order magnetic wakefield is given by:

$$B_{\theta}^{(1)} = \frac{r m_e c^2}{2 e r_0^2}\left(\sum_{s=1}^{2} a_{rs}^2 \Pi_s^2 \omega_p^2 \frac{\omega_s^4+\omega_s^2\omega_c^2}{\left(\omega_s^2-\omega_c^2+i\nu_c\omega_s\right)^2}\right)\left[1-Cos\frac{2\pi\xi}{L}\right] \tag{17}$$

Proceeding with the second iteration, the respective second-order longitudinal and transverse wakefields are obtained as follows:

$$E_z^{(2)} = \frac{m_e k_p^2 c^2 n_{Pe}}{16 e n_u}\left(\sum_{s=1}^{2}\frac{\mathrm{a}_{rs}^2\Pi_s^2\omega_p^2}{\left(1-\frac{k_p^2L_s^2}{4\pi^2}\right)}\frac{\omega_s^4+\omega_s^2\omega_c^2}{(\omega_s^2-\omega_c^2+i\nu_c\omega_s)^2}\right)$$

$$\times\Big\{Sink_pLCosk_p\xi + L\left[Cosk_p\xi - Cosk_p(L-\xi)\right] - \tfrac{1}{2}\left[Sink_p(2L-\xi)+Sink_p\xi\right] - \left(2+\frac{1}{k_p^2r_0^2}\left(1-\frac{2r^2}{r_0^2}\right)\right)\left[Sink_p(L-\xi)+Sink_p\xi\right]\Big\} \tag{18a}$$

$$E_r^{(2)} = \frac{m_e c^2}{4e}\left(\sum_{s=1}^{2}\frac{\mathrm{a}_{rs}^2\Pi_s^2\omega_p^2 r}{\left(1-\frac{k_p^2L_s^2}{4\pi^2}\right)}\frac{\omega_s^4+\omega_s^2\omega_c^2}{(\omega_s^2-\omega_c^2+i\nu_c\omega_s)^2}\right)$$

$$\Big\{-\left(2+\frac{16n_u}{n_{Pe}k_p^2r_0^2}\left(1-\frac{r^2}{r_0^2}\right)\right)\left[Cosk_p\xi - Cosk_p(L-\xi)\right] + k_pL\left[Sink_p\xi + Sink_p(L-\xi)\right] + Sink_pLSink_p\xi + \left(\sum_{s=1}^{2} 1-\frac{k_p^2L_s^2}{4\pi^2}\right)Sin^2\frac{\pi\xi}{L}\Big\} \tag{18b}$$

and the second-order of magnetic wakefield is expressed as:

$$B_{\theta}^{(2)} = \frac{m_e c^2}{4e}\left(\sum_{s=1}^{2}\mathrm{a}_{rs}^2\Pi_s^2\omega_p^2\frac{\omega_s^4+\omega_s^2\omega_c^2}{(\omega_s^2-\omega_c^2+i\nu_c\omega_s)^2}\right)$$

$$\times\left\{\frac{k_pLn_{Pe}}{n_u}\left(\sum_{s=1}^{2}\frac{r}{\left(1-\frac{k_p^2L_s^2}{4\pi^2}\right)}\right)\left[Sink_p(L-\xi)+Sink_p\xi\right] + \left(1-Cos\frac{2\pi\xi}{L}\right)\right\} \tag{19}$$

When a short, intense laser pulse propagates through plasma, it excites plasma waves via the ponderomotive force. Electrons can become trapped in these waves and accelerated by their electric fields, thereby gaining energy from the wake. The electron energy gain in laser–plasma wake-field acceleration quantifies the amount of energy electrons acquire as they ride the wake created by a high-intensity laser pulse. The normalized energy gain associated with the change in the relativistic factor is expressed as:

$$\frac{\Delta W}{E_{rest}} = \frac{e}{m_ec^2(1-\beta)}\int\left(E_z^{(1)}+E_z^{(2)}\right)d\xi \tag{20}$$

Where $E_{rest}$ is the electron rest energy. This relation underscores the crucial role of wake-field dynamics in THz generation, providing a basis for optimization. Specifically, stronger plasma electron oscillations correspond to higher output power and broader bandwidth, particularly in magnetized plasma. Moreover, by tailoring laser and plasma parameters, one can achieve precise control over the characteristics of THz waves, including their power, frequency, coherence, and spectral tunability.

The generated longitudinal ($E_z$) and radial ($E_r$) wakefields, along with the associated azimuthal magnetic field ($B_\theta$), drive nonlinear plasma currents that serve as sources for THz electromagnetic radiation. In the presence of the external axial magnetic field $B_0$, the transverse (radial) component of the wakefield induces a significant nonlinear transverse current density $J_r$ and/or $J_\theta$, modulated at frequencies near the plasma frequency $\omega_p$ or its harmonics. This current, combined with the rippled density profile facilitating phase matching, acts as a radiating antenna according to the inhomogeneous wave equation derived from Maxwell's equations. The radiated THz field in the far field arises primarily from the transverse oscillations, with the magnetic field enhancing coherence and directing emission patterns (often conical or radially outward). Resonant coupling between wakefield harmonics and the chirp-modulated laser beat frequency further amplifies specific THz peaks, as observed in the Fourier spectra. While a full analytical solution for the radiated power spectrum is beyond the quasi-static fluid approximation here, the wakefield amplitudes derived in Eqs. (15)–(19) provide the driving source strength for THz emission, consistent with the FBPIC simulation results presented in next Section. he radiated THz spectra are thus expected to reflect the Fourier content of these nonlinear currents, with angular distributions shaped by the cylindrical symmetry and magnetic field, as quantified in the FBPIC simulations via far-field field extraction. Quantitative validation of the theoretical wakefield amplitudes and their parametric dependence (chirp $\alpha$, density $n_0$, $B_0$) is provided by direct comparison with high-resolution FBPIC simulations in Section III, where spatiotemporal field profiles and Fourier spectra confirm the predicted enhancements in wakefield strength and resulting THz peak amplitudes.

## III. Results and Discussion

The theoretical model in Section II establishes the analytical foundation for wake-field excitation, examining key parameters individually for clarity. The FBPIC simulations build upon this framework by capturing their combined nonlinear interactions and the resulting enhancement of THz emission from wake-field-driven currents. The results of this study show that copropagating chirped laser pulses in a rippled, magnetized plasma channel can drive both longitudinal and transverse wake-fields whose characteristics depend sensitively on laser chirp, pulse duration, and the background plasma profile. As illustrated schematically in Figure 1, the mechanism originates from the beat-frequency modulation of the ponderomotive force produced when two chirped pulses, with distinct angular frequencies and wave numbers, propagate along the plasma axis. The schematic depicts the lasers co-propagating along the longitudinal axis (z-direction) of the cylindrical plasma, with an external axial magnetic field $B_0$ applied parallel to the propagation direction to confine electron motion, while THz emission arises radially from the nonlinear wakefield currents. The combined action of this force and the externally applied magnetic field initiates electron oscillations both axially and transversely. These oscillations create charge separation and give rise to both longitudinal and transverse wakefields, governed by the fluid-Maxwell system formulated in Eqs. (10)–(14). The visualized cylindrical plasma channel in the schematic captures the axisymmetric assumption underlying the theoretical model, where radial variations are represented with a rippled density. The resulting wakefields, are nonlinear structures whose formation and evolution depend sensitively on laser chirp, pulse duration, and the background plasma profile.

The generation of radial and longitudinal wake-fields by two co-propagating chirped laser pulses in a magnetized plasma was further examined using Fourier-Bessel Particle-In-Cell (FBPIC) simulations. This quasi-three-dimensional simulation framework, optimized for cylindrical geometries, combines the efficiency of azimuthal Fourier decomposition with the accuracy of full PIC methods, enabling realistic modeling of relativistic plasma dynamics at reduced computational cost. In this scheme, the plasma is represented by macroparticles whose motion follows the relativistic Lorentz force equation. The electromagnetic fields are discretized on a cylindrical mesh and evolved self-consistently using Maxwell's equations. The computational domain was configured as a cylindrical volume with a moving simulation window of length $10\lambda_p$ along the propagation axis $z$, where $\lambda_p = 2\pi c/\omega_p$ is the plasma wavelength determined by the plasma frequency $\omega_p$. Perfectly Matched Layer (PML) boundary conditions were applied along the longitudinal direction to absorb outgoing waves and eliminate non-physical reflections. The spatial resolution was set to $\Delta r \approx \lambda_p/20$ in the radial direction and $\Delta z \approx c\Delta t$ in the longitudinal direction, with the time step $\Delta t$ chosen to satisfy the Courant–Friedrichs–Lewy (CFL) stability condition. This configuration ensures accurate capture of the wakefield structures and preserves numerical stability during the simulation. The results confirm that the interplay of beat-frequency modulation and the external magnetic field produces both axial and transverse electron oscillations, consistent with the fluid-Maxwell model, while also revealing how chirp and plasma inhomogeneity shape the strength and morphology of the wake-fields.

To perform the FBPIC simulations and validate the theoretical model, the following parameters are used: The two co-propagating chirped laser pulses have central wavelengths $\lambda_1 = 800nm$ and $\lambda_1 = 805nm$ (corresponding to central frequencies $\omega_1 = 2.36 \times 10^{15}\, rad/s$ and $\omega_2 = 2.34 \times 10^{15}\, rad/s$), such that the beat frequency $\Delta\omega \approx \omega_p$. Each pulse has a Gaussian temporal envelope with FWHM duration $\tau = 30 - 50fs$, Gaussian radial profile with spot size $w_0 = 10 - 20\mu m$, and normalized vector potential $a_0 = eA/m_e c \approx 1 - 3$ (peak intensity $I = 10^{18} - 10^{19}\, W/cm^2$ for typical Ti:sapphire lasers). The pulses are focused on-axis at the plasma entrance and co-propagate along the z-direction in the moving simulation window. The background plasma is underdense with unperturbed electron density $n_0 = 10^{19} cm^{-3}$, corresponding to plasma frequency $\omega_p = 5.6 \times 10^{12}\, rad/s$ (~0.9 THz). An external axial magnetic field $B_0 = 1 - 10T$ is applied along $z$. The plasma is modeled as rippled with modulation amplitude ~10–20% and period matched to wake/resonance conditions. The simulation domain uses a cylindrical moving window with high resolution ($\Delta z \approx \lambda_p/100$, ($\Delta z \approx w_0/20\Delta r \approx w_0/20$) to resolve wake structures and outgoing THz waves.

The 3D visualizations shown in Figure 2 further illustrate the spatial and temporal evolution of the wakefields generated by the interaction of chirped laser pulses with a rippled-magnetized plasma. These results highlight the multidimensional nature of the plasma's nonlinear response. The longitudinal wakefield, associated with the axial electric field component, displays a dominant central peak aligned along the normalized time and phase coordinates, reflecting the influence of the ponderomotive force in driving electron oscillations along the propagation axis. The spatial coherence of this field, stabilized by the externally applied magnetic field, aligns with theoretical predictions based on the coupling of Maxwell's equations with the fluid momentum equations. In contrast, the interplay between the laser chirp, transverse intensity gradients, and plasma response leads to the emergence of field patterns, which are sensitive to the temporal and spatial chirp parameters. The radial wakefield exhibits a more intricate structure with multiple lobes, stemming from the transverse ponderomotive force induced by the Gaussian intensity profile of the laser pulses. These lobes represent alternating regions of electron depletion and accumulation, indicating the formation of transverse charge separation.

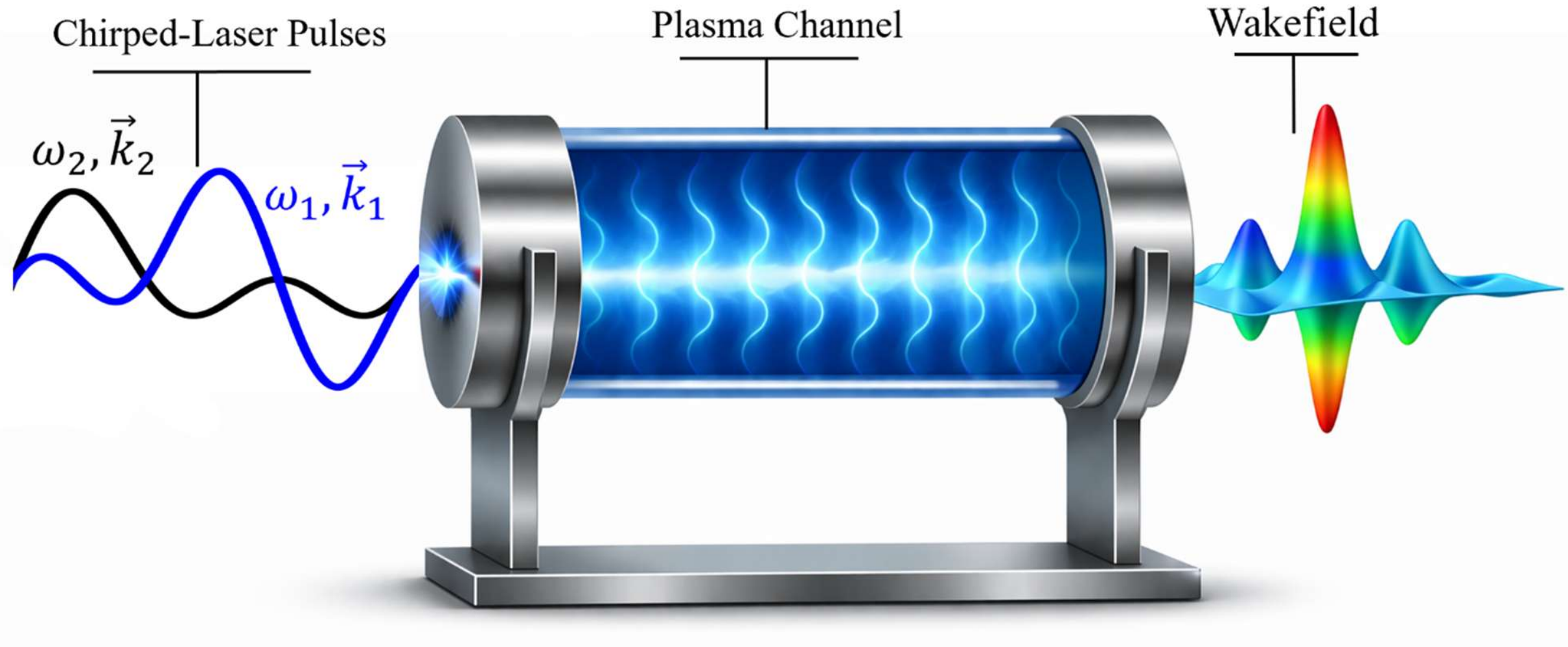

**Fig. 1.** Schematic illustration of two co-propagating chirped laser pulses propagating along the longitudinal ($z$) axis of a cylindrical rippled-density magnetized plasma channel. The external static magnetic field $B_0$ is applied along the propagation direction ($z$-axis), and THz radiation is emitted radially/transversely from the wakefield-driven currents.

Figures 3 illustrates the spatiotemporal evolution of longitudinal ($E_z$) and transverse ($E_r$) fields for various plasma densities. The dynamics of these fields are governed by the combined effects of the laser-driven ponderomotive forces, external magnetic field, and the anisotropic dielectric response of the plasma, all modulated by the unperturbed electron density. The longitudinal field reveals a directionally confined angular emission pattern shaped by magnetic confinement and longitudinal charge separation. Its temporal evolution shows periodic modulations that become increasingly pronounced at higher electron density, reflecting a stronger ponderomotive coupling and enhanced momentum transfer. This leads to deeper electron oscillations and more efficient wakefield generation. Spatially, $E_z$ demonstrates sharper oscillatory structures and increased amplitude at higher densities, indicating steeper potential gradients and stronger longitudinal wakefields. The attenuation of high-frequency components suggests energy dissipation mechanisms such as collisional damping and finite pulse duration, incorporated in the model via a density-dependent collision frequency and pulse envelope profile. Complementarily, the transverse field captures the radial wakefield dynamics resulting from laser–plasma interaction in cylindrical symmetry. Its angular and temporal patterns display oscillations with density-dependent amplitude and phase modulation. As the electron density increases, the transverse ponderomotive force—originating from the non-uniform radial laser intensity—induces stronger radial electron displacement, giving rise to amplified transverse fields. The spatial profile of $E_r$, plotted versus normalized wave number, shows a periodic pattern with enhanced modulation depth at higher densities, reflecting the nonlinear amplification of radial charge separation and the constructive buildup of transverse wake-fields. Similar to the longitudinal case, the decline in amplitude at higher wave numbers is attributed to dissipative effects modeled through collisional terms and pulse-shape constraints. The resulting THz spectrum, characterized by distinct peaks, is consistent with experimental observations, reinforcing the connection between plasma response, wake-field dynamics, and radiation output [34, 35].

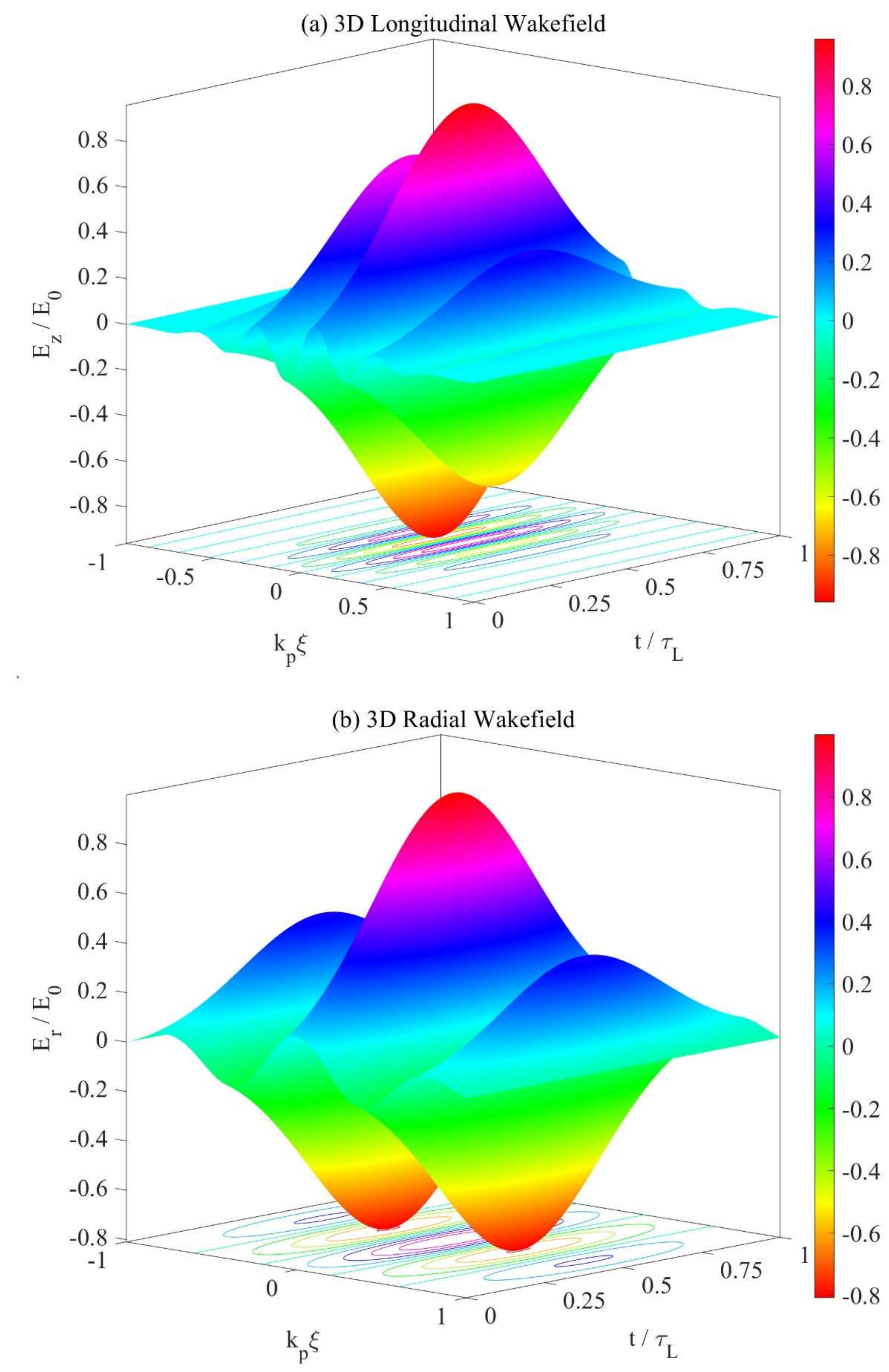

**Fig. 2.** The spatial and temporal evolution of the longitudinal and radial wakefields with their contours.

The analysis of the influence of the laser frequency chirp parameter on both the longitudinal and radial electric fields is presented in Figure 4. The results show that the temporal and angular evolution of $E_z$ and $E_r$ undergo distinct chirp-dependent phase shifts and amplitude modulations. These modulations alter the ponderomotive force acting on plasma electrons, which in turn modifies their oscillatory dynamics and the efficiency of wakefield excitation. As $\alpha$ increases—particularly for positive chirps—both longitudinal and transverse field components display enhanced initial peak amplitudes, indicating stronger charge separation and more efficient energy deposition. The spatial profiles of $E_z$ and $E_r$ further demonstrate how $\alpha$ controls the periodicity, amplitude, and decay of the wakefields along the laser propagation axis. In both field components, larger $\alpha$ leads to sharper oscillations and stronger field amplitudes, reflecting intensified laser–plasma coupling. The observed attenuation at higher wave numbers is attributed to dissipative mechanisms, including collisional damping and finite pulse effects, whose impact is modulated by the chirp profile of the laser pulse. Beyond these one-dimensional profiles, Figure 5 provides a 2D spatiotemporal characterization of the longitudinal and radial wake-fields driven by chirped laser pulses in rippled-magnetized plasma. The radial wakefield depends on chirp, with its amplitude and spatial distribution shaped by the transverse ponderomotive force. The longitudinal wake-field shows pronounced structural modulation across normalized time and plasma wave number. For negative chirp, field intensity varies asymmetrically across specific phases, reflecting accelerated electron oscillations driven by the temporal frequency shift. In contrast, zero and positive chirp, yield more symmetric distributions, suggesting a balanced laser-plasma interaction modulated by the external magnetic field. Positive chirp also enhances longitudinal wake-field amplitude, in agreement with experimental observations that Gaussian chirped pulses significantly increase wake-field strength [36].

The spatiotemporal variations of the longitudinal and radial wake-fields, along with their corresponding radiation patterns, are shown in Figure 6. The influence of the external magnetic field is evident in both the spatial distribution and temporal evolution of the wakefields. By introducing anisotropy into the plasma response, the magnetic field modifies the dispersion relations of plasma modes and generates distinct angular radiation patterns. The longitudinal electric field displays oscillatory structures whose angular spread and amplitude increase with pulse duration, suggesting that the combined effect of the pulse envelope and magnetic confinement enhances energy coupling into specific plasma modes. The magnetic field also suppresses transverse electron motion, redirecting more of the laser's energy into longitudinal modes and thereby reinforcing the axial wakefield. In addition, Figure 7 illustrates the spatiotemporal evolution of wakefields generated by chirped laser pulses in a magnetized plasma, highlighting how varying magnetic field strength reshapes their structures. Both field components show that wake-field amplitude and localization are highly sensitive to the applied field. Stronger magnetic fields enhance laser-plasma coupling and reduce spatial spread focusing the energy transfer and reinforcing the wakefield intensity. The longitudinal field becomes more confined and amplified, a result of the increased Lorentz force that alters electron trajectories and strengthens the axial ponderomotive drive. Similarly, the radial wake-field sharpens with increasing $B_0$ as the magnetic field constrains transverse electron motion, resulting in more localized radial field structures. These results are consistent with theoretical predictions that the external magnetic field modifies both dispersion relations and nonlinear plasma response, and the observed enhancement of wake-field intensity with increasing magnetic field strength aligns with experimental findings [37, 38]. These figures demonstrate the parametric dependence of wakefield amplitude and structure on density, chirp, and magnetic field, revealing enhanced energy transfer and coherence for THz generation through resonant ponderomotive driving.

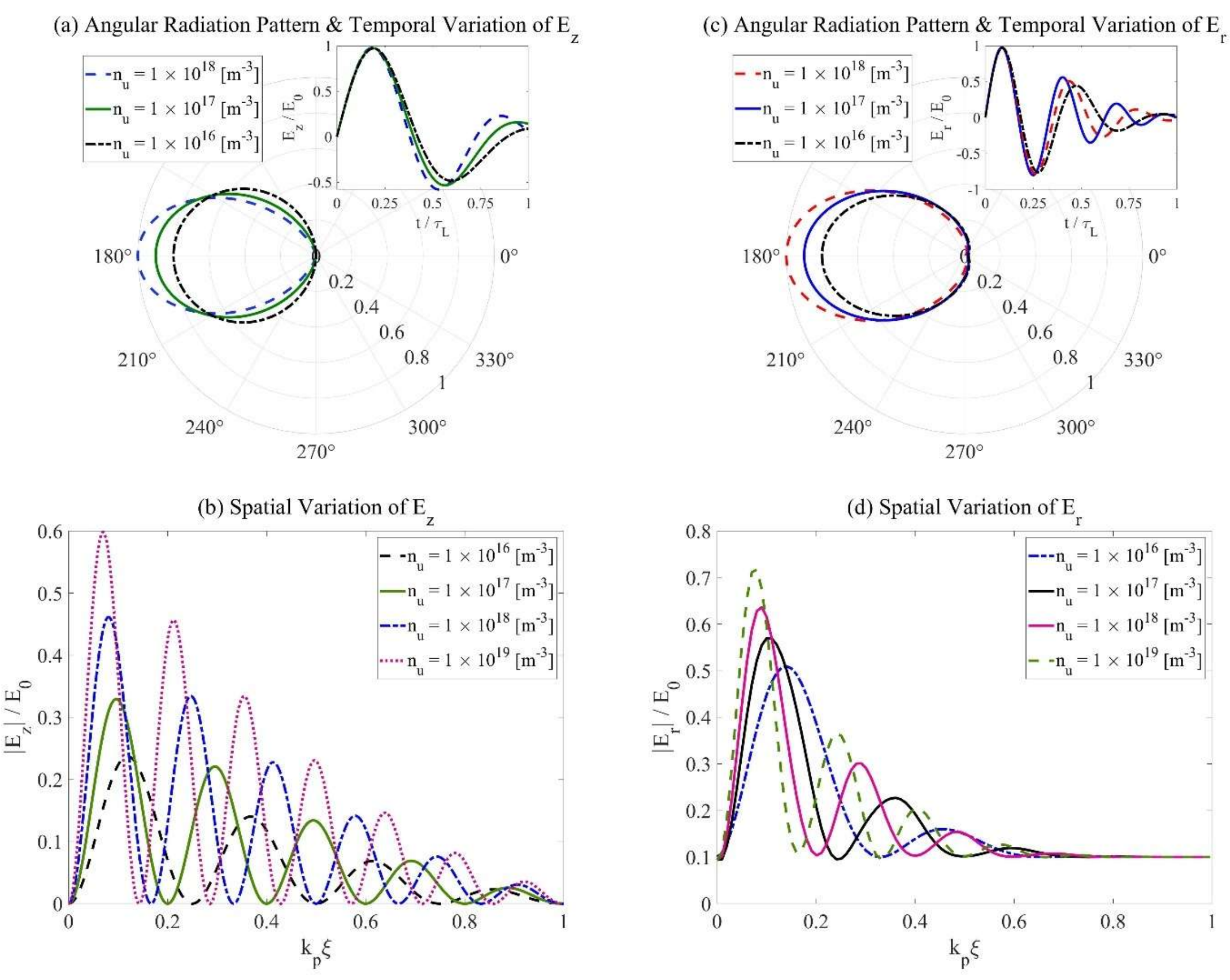

**Fig. 3.** The spatiotemporal variations of the radial and longitudinal wakefield for various plasma densities.

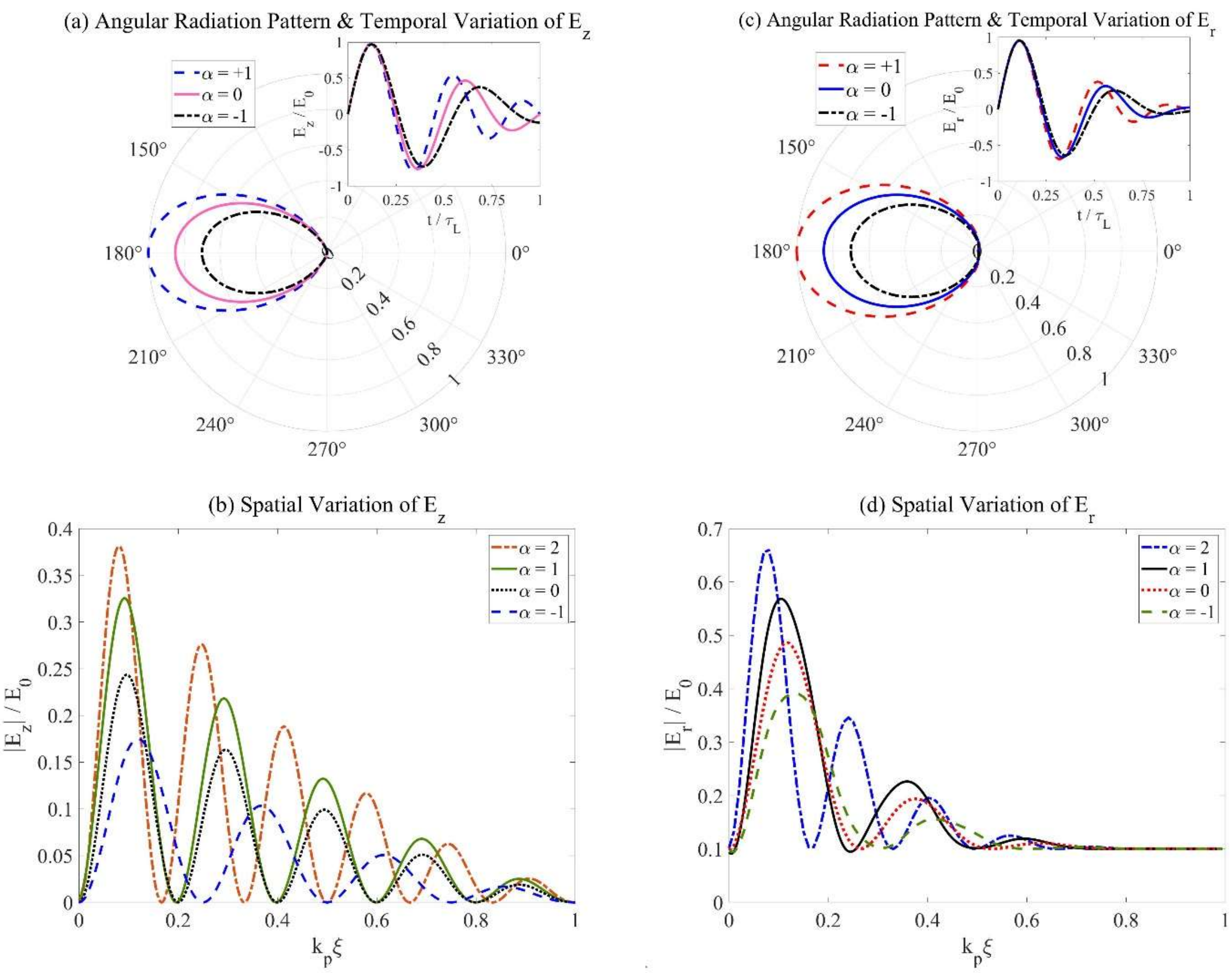

**Fig. 4.** Role of chirped parameter on the spatiotemporal variations of the radial and longitudinal wakefield.

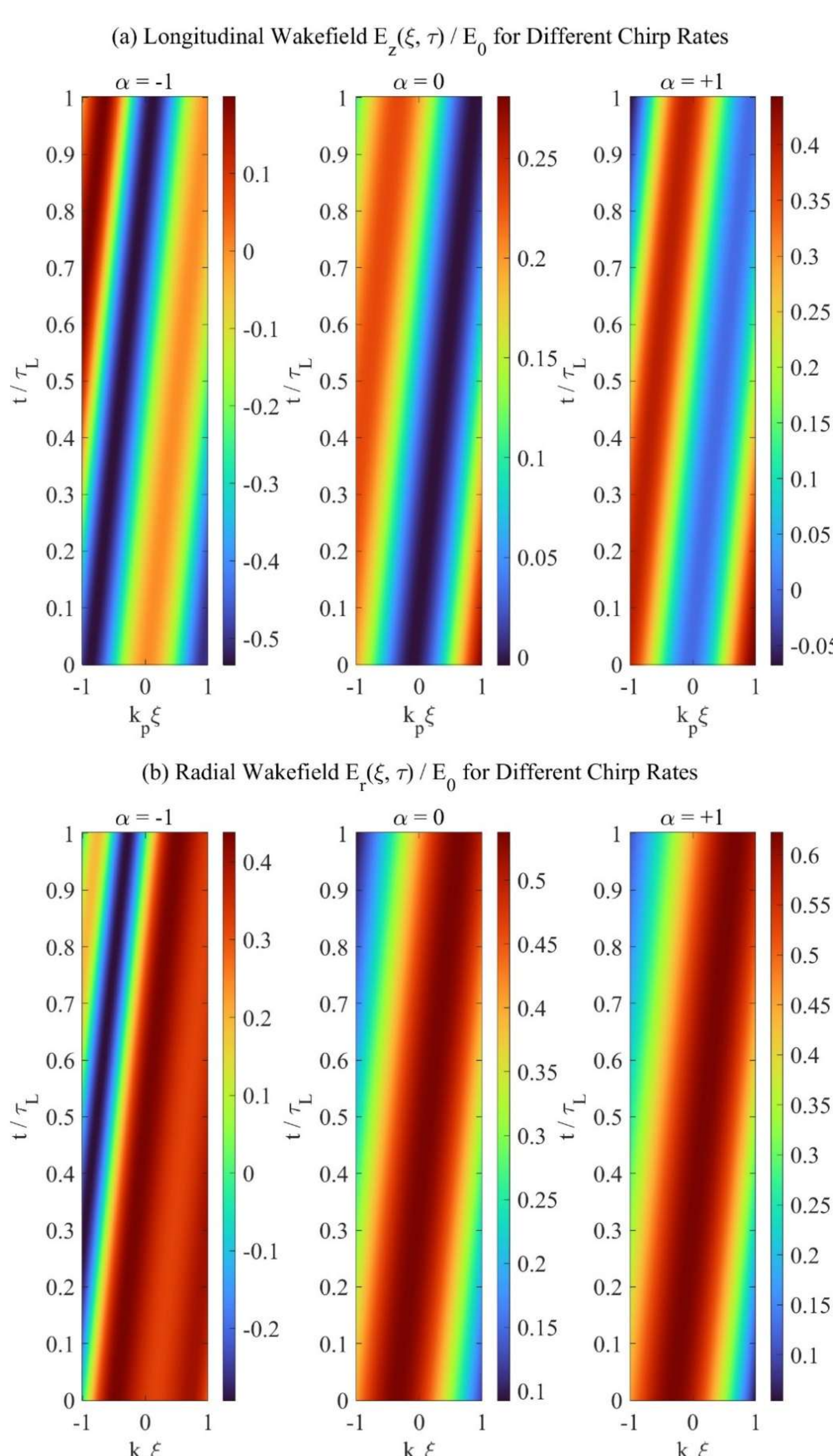

**Fig. 5.** Impact of chirped parameter on the spatiotemporal variations of the longitudinal and radial wakefields.

The role of laser pulse length in shaping the spatial and temporal variations of longitudinal and radial wakefields, as well as the corresponding radiation patterns, is illustrated in Figure 8. As the pulse length increases, the wake-fields exhibit more sustained and oscillatory temporal evolution, accompanied by a broader angular spread of the radiation pattern. This behavior indicates that a longer interaction times between the laser pulse and plasma enhance the ponderomotive force, leading to stronger electron displacement and wider emission profile under the influence of the external magnetic field. In contrast, shorter pulses lead to more localized field structures in both time and space, reflecting a reduced interaction interval and a more confined energy deposition. Along the propagation axis, longer pulses produce higher initial field amplitudes and more developed oscillatory patterns, suggesting a larger interaction volume that facilitates greater charge separation and stronger wake excitation in both longitudinal and transverse components. Shorter pulses, however, generate weaker fields that decay more rapidly, consistent with the limited spatial extent and reduced energy transfer. These trends suggest that by optimizing the temporal profile of the pulse, one can effectively control the strength, persistence, and angular distribution of the induced fields, thereby enhancing mechanisms relevant to plasma-based electron acceleration and THz radiation generation.

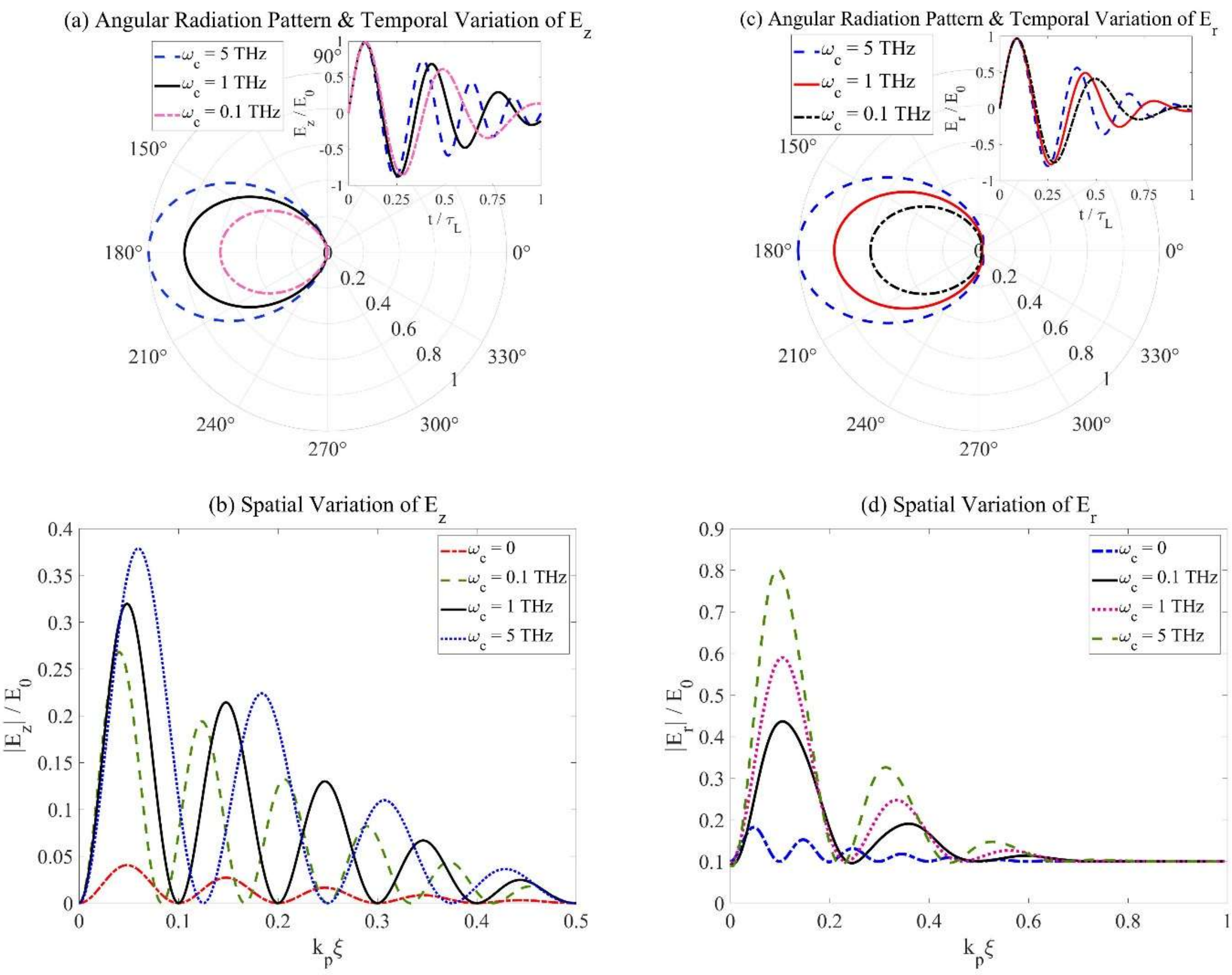

**Fig. 6.** The spatiotemporal variations of the longitudinal and radial wakefield for various magnetic field.

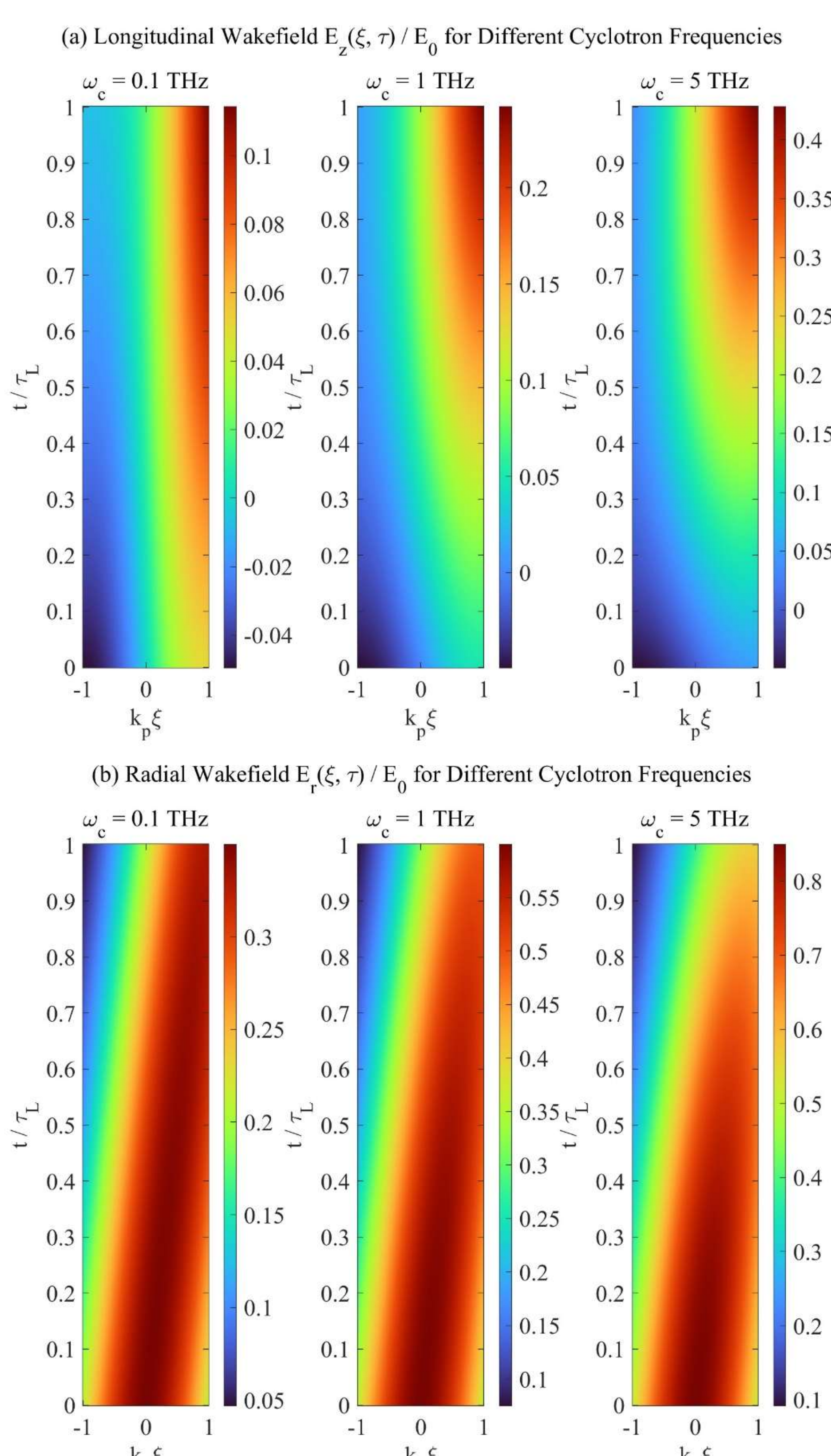

**Fig. 7.** Impact of magnetic field on the spatiotemporal variations of the longitudinal and radial wakefields.

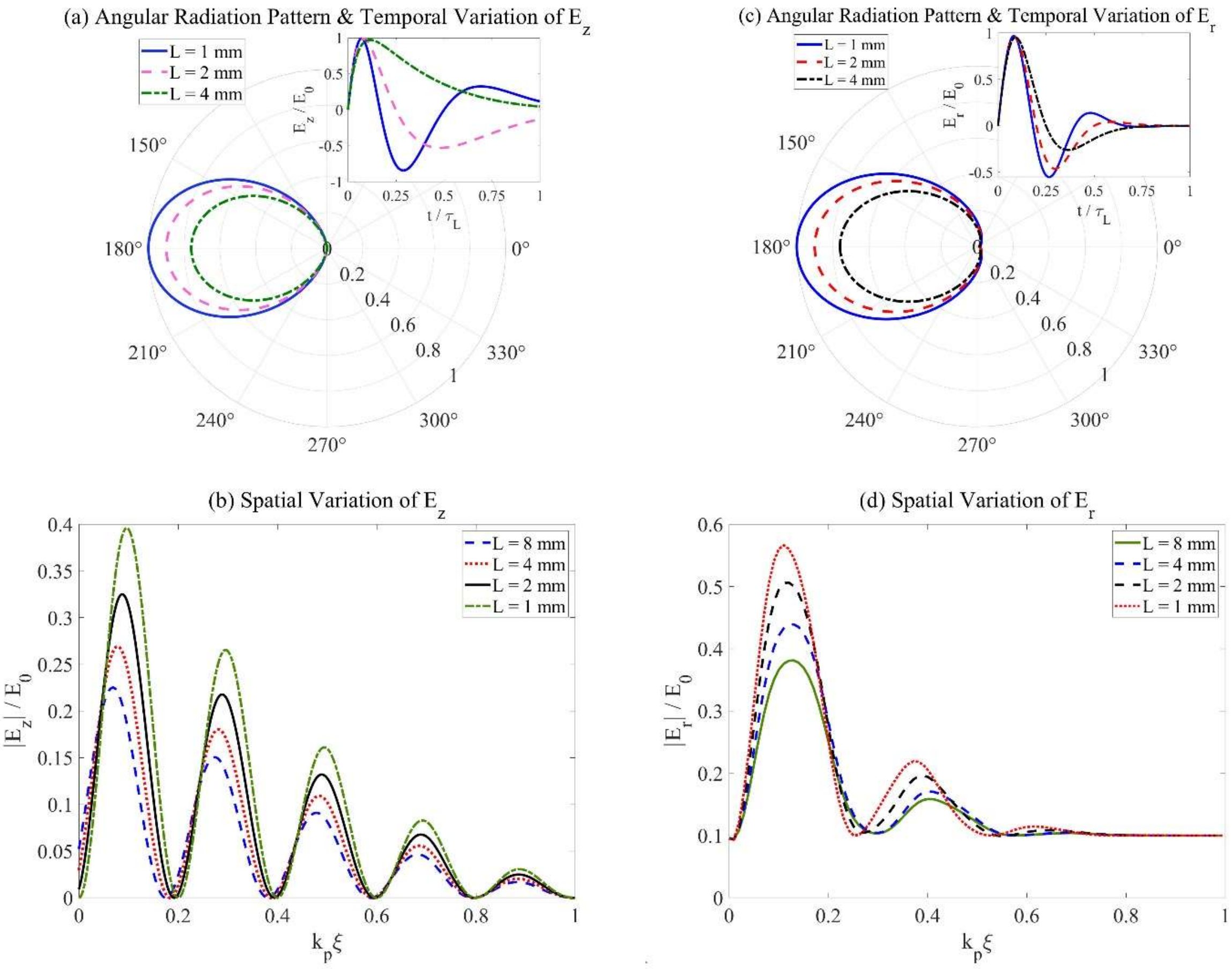

**Fig. 8.** Role of laser pulse length on the spatiotemporal variations of the radial and longitudinal wakefield.

The variations in normalized electron energy gain wakefields generated behind the laser pulses with different chirp parameters are shown in Fig. 9a. As the chirp parameter increases, the electron energy gain is enhanced. This behavior arises because changing the pulse frequency modifies the laser–plasma resonance condition. At lower frequencies, the resonance with plasma oscillations is less effective, limiting energy transfer. As the pulse frequency increases, it more closely synchronizes with plasma oscillations, thereby improving coupling efficiency. The higher frequency also extends the effective interaction time during which the laser pulse can sustain strong wake-fields, resulting in prolonged energy transfer and higher electron acceleration. Figure 9b illustrates the effect of the cyclotron frequency on the normalized electron energy gain. The external magnetic field confines electron motion by forcing electrons to gyrate around the magnetic field lines, which prevents radial spreading and yields more directed dynamics. The magnetic field confinement not only restricts transverse motion but also enhance longitudinal coherence between plasma electrons and the driving laser pulse. Stronger coherence improves the efficiency of energy transfer from the wakefield to plasma electrons, leading to greater acceleration along the propagation axis. The accompanying heatmaps illustrate the spatial distribution of normalized electron energy gain across the radial (r) and axial (z) coordinates. In Fig. 9a, the heatmap highlights how the wakefield intensity evolves with the chirp parameter (α), showing a more pronounced and structured energy distribution. In Fig. 9b, the confinement effect of the magnetic field is evident: high-energy regions become concentrated along the propagation axis, consistent with the enhanced longitudinal coherence observed in the line plots. Together, these heatmaps highlight the spatial dynamics underlying the increased electron acceleration. The resulting enhancement of electron energy gain aligns well with both numerical predictions and experimental findings [39, 40]. The spatial and temporal correlations of two laser pulses, as a function of their propagation and interaction dynamics, is illustrated in Fig. 10. The spatial correlation quantifies the degree of transverse overlap between the intensity profiles of the two chirped laser pulses at each propagation distance within the plasma. It is evaluated by comparing the radial intensity distributions extracted directly from the simulated electric field data in the FBPIC moving window, starting near perfect alignment at the plasma entrance and decreasing over distance due to effects such as differential self-focusing, defocusing, collisional scattering, and plasma-induced distortions. The temporal correlation assesses the synchronization and phase coherence between the two pulses along the propagation axis. It is characterized by analyzing the temporal envelopes or on-axis field oscillations at successive positions (or time steps in the simulation frame), initially showing mismatch, then reaching a maximum at intermediate propagation distances through plasma-mediated mechanisms including cross-phase modulation and group velocity matching, before eventual decay from dephasing, dispersion, and dissipative processes. At the plasma entrance, the pulses exhibit a high degree of spatial correlation, indicating strong overlap in their transverse spatial profiles. As they propagate. however, this spatial correlation diminishes due to collisional effects and nonlinear interactions, that disrupt the transverse alignment. The density gradients in the plasma, together with relativistic self-focusing or defocusing, further contribute to spatial decorrelation, leading to increasingly randomized transverse distribution over time. In contrast, the temporal correlation follows a different trajectory. Initially, it is relatively low, reflecting a phase mismatch between the two pulses at the start of their interaction. As the pulses propagate, the temporal correlation increases, reaching a maximum at an intermediate stage. This peak arises from the plasma's nonlinear response, which enables the pulses to synchronize their phases through mechanisms such as cross-phase modulation and plasma-mediated group velocity matching. The maximum temporal correlation signifies an optimal condition for energy transfer, where the combined electric fields reinforce each other, amplifying the wakefield amplitude and enhancing electron acceleration efficiency. Subsequently, the temporal correlation decreases as accumulated phase distortions, dispersive effects, collisional dissipation, and the finite pulse duration degrade the synchronized state.

The normalized intensity spectra -- obtained from the Fourier transform of the electric field during laser-plasma interaction and plotted as a function of frequency in the THz range – exhibit distinct peaks decoupled from the plasma response. The THz radiation spectra and angular patterns are extracted from the Fourier transform of the electromagnetic fields recorded at virtual probes/detectors placed in the far-field region or outside the plasma (using the simulation's PML boundaries to allow outgoing waves), capturing the propagating radiated components rather than the near-zone wakefields. This approach accounts for the nonlinear transverse currents driven by radial wakefields and magnetic confinement, which act as sources for THz emission according to Maxwell's equations. The resulting spectra show distinct peaks in the THz range, modulated by pulse duration, laser intensity, plasma density, and chirp, with resonant enhancement when wakefield harmonics align with the modulated laser beat frequency. The positions and amplitudes of these peaks are modulated by parameters such as pulse duration, laser intensity, and plasma density, reflecting the underlying nonlinear plasma dynamics. The THz emission originates from nonlinear currents and transverse electric field components driven by electron oscillations within the wake-fields, as evidenced by the Fourier spectra shown in Figure 11. These nonlinear currents arise from the collective motion of plasma electrons displaced by the laser's ponderomotive force, which expels electrons from high-intensity regions and produces time-varying charge densities that act as electromagnetic sources in accordance with Maxwell's equations. In the cylindrical geometry of the rippled magnetized plasma, radial electron displacements couple with azimuthal magnetic fields to generate transverse propagating modes that radiate as free-space THz waves – consistent with the 2D wake-field dynamics reproduced in FBPIC simulations. Under resonant conditions, when the plasma frequency or wakefield harmonics align with the modulated frequency of the chirped laser pulses, the peak amplitudes are further enhanced through efficient energy transfer from the laser to the plasma modes. The high-energy THz emission observed in the spectrum is consistent with experimental observations [41, 42].

The proposed scenario of two co-propagating chirped laser pulses interacting in a rippled, magnetized plasma channel is experimentally realizable with current high-power laser facilities. Chirped pulses can be generated and synchronized using standard chirped pulse amplification (CPA) systems, with beat-frequency control achieved by splitting a single chirped beam or employing optical parametric amplification for independent frequency/chirp tuning. The rippled density profile can be produced via pre-ionization with a modulated laser beam, periodic capillary structures, or engineered gas nozzles to create density gratings. An external axial magnetic field (several Tesla) can be applied using solenoid coils surrounding the plasma target, as routinely demonstrated in magnetized laser wakefield experiments to confine electron motion and enhance wake coherence. Underdense plasma targets enable the required relativistic intensities ($a_0 \sim 1-5$) with terawatt-class lasers. THz emission can be diagnosed using established far-field electro-optic sampling or bolometric methods. While combining all elements (chirped beat-wave, rippled density, axial magnetization) in one experiment remains unexplored, individual components have been verified in related studies on beat-wave plasma acceleration, magnetized THz generation, and ripple-enhanced radiation, suggesting high potential for experimental validation and optimization of the predicted wakefield-driven THz enhancements.

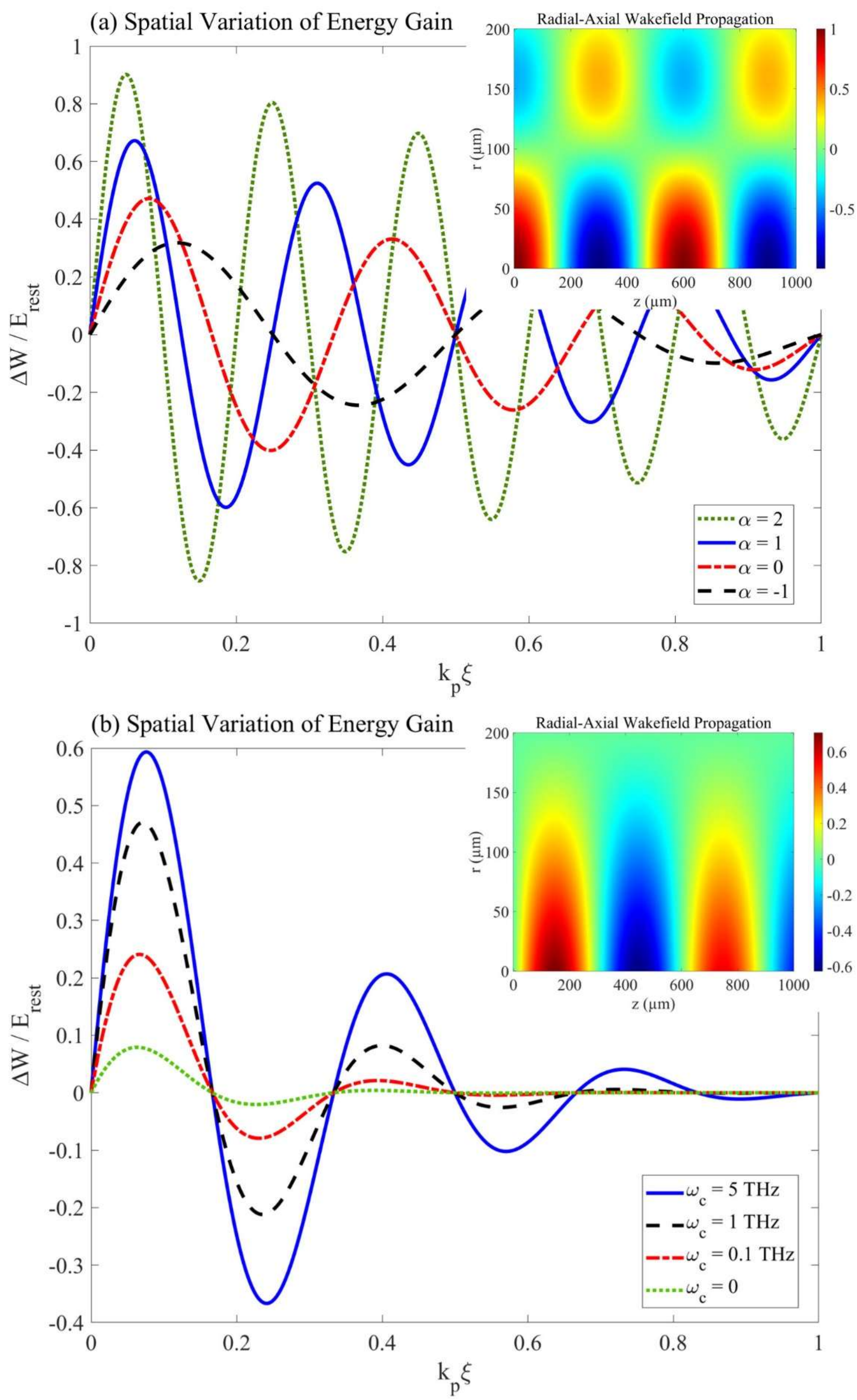

**Fig. 9.** The spatial variations of energy gain for various chirped parameter and cyclotron frequencies.

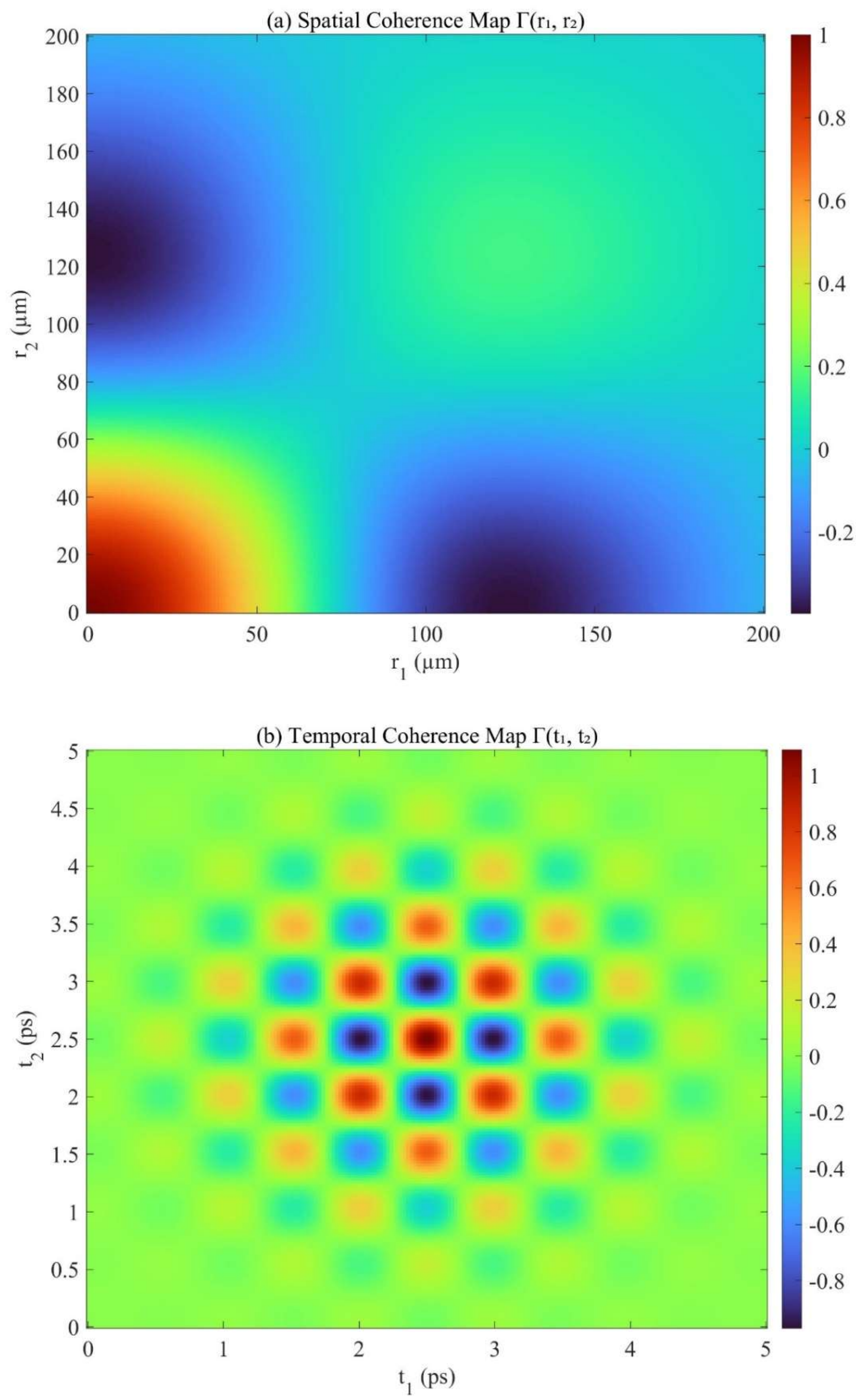

**Fig. 10.** Evolution of spatial overlap and temporal synchronization between the two co-propagating chirped laser pulses as functions of normalized propagation distance in the rippled magnetized plasma. The spatial correlation reflects transverse intensity alignment, while temporal correlation indicates phase and envelope coherence, both influenced by plasma nonlinearities and dissipation.

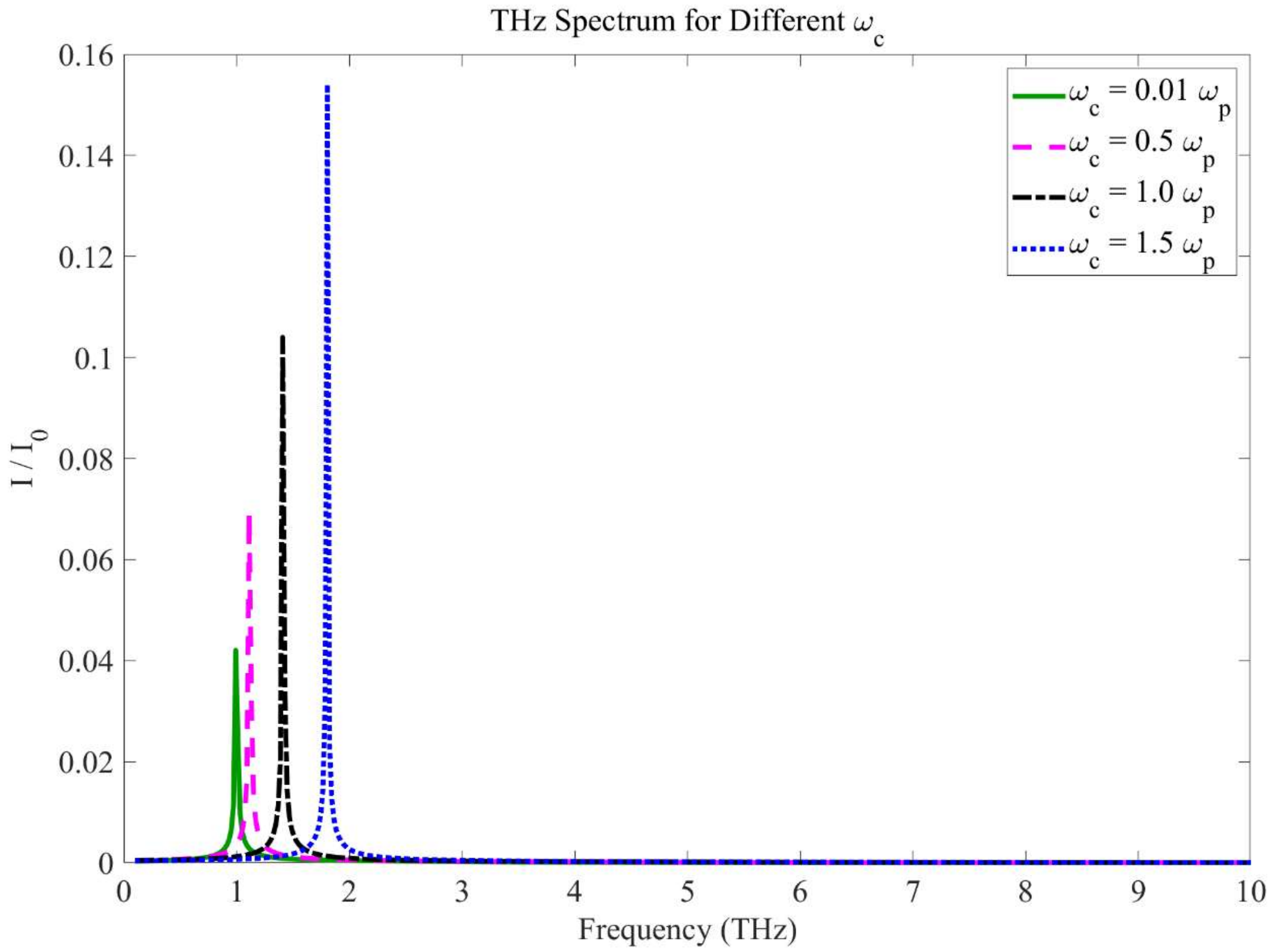

**Fig. 11.** Normalized intensity spectra in the THz range obtained from the Fourier transform of the propagating electromagnetic fields (far-field components) extracted from FBPIC simulations of the chirped-laser interaction with rippled-magnetized plasma, showing distinct resonant peaks modulated by laser and plasma parameters.

## IV. Conclusions

This study has characterized the spatiotemporal evolution of radial and longitudinal wakefields generated by two co-propagating chirped laser pulses in a magnetized plasma channel. It be shown that the combination of a fluid-Maxwell analytical model and high-resolution Fourier-Bessel Particle-In-Cell (FBPIC) simulations in cylindrical geometry gives us more control over wakefield formation, electron energy gain, and THz radiation characteristics than ever before. This is because of the interaction of laser chirp, pulse duration, plasma density modulation, and external magnetic field strength.

The findings demonstrate that positive laser chirps markedly increase wakefield amplitudes and coherence by dynamically altering the beat frequency and enhancing resonance with plasma oscillations via the ponderomotive force. The external axial magnetic field is very important for keeping transverse electron motion in check, reducing instabilities, and strengthening longitudinal phase coherence. This, in turn, leads to stronger axial fields and a much higher normalized electron energy gain. Also, the rippled density profile helps with better phases matching and keeps wake structures going over long distances. Initially, there are a lot of spatial overlaps between the two pulses at the entrance to the plasma, but this overlap fades over distance because of collisional scattering, nonlinear distortions, and dispersive effects. Actually, temporal synchronization reaches its highest point in the middle of the plasma through plasma-mediated cross-phase modulation and group velocity matching, before eventually falling off due to accumulated phase distortions and dissipation. Fourier analysis of the propagating electromagnetic fields extracted from the simulations confirms the generation of strong, tunable THz radiation arising primarily from nonlinear transverse currents driven by radial wakefields and enhanced by magnetic confinement, with peak intensities and spectral features highly sensitive to the explored laser and plasma parameters. These findings provide new insight into optimizing chirped-pulse laser–plasma interactions, offering pathways toward controlled wake-field acceleration and the development of compact, tunable THz sources.

**Acknowledgment**

This work did not receive any specific grant from funding agencies in the public, commercial, or not-for-profit sectors.

**Author Contributions**

Ali Asghar Molavi Choobini was responsible for conceptualization, data curation, formal analysis, investigation, methodology, writing the original draft, and review and editing, all with equal contributions. Farzin M. Aghamir contributed equally to investigation, project administration, supervision, validation, and review and editing of the manuscript.

**Data availability statement**

The data that support the findings of this study are available from the corresponding author upon reasonable request.

**Competing interests**

The authors declare no competing interests.

**Funding declaration**

The authors received no financial support for the research, authorship, and/or publication of this article.